\documentclass[preprint, aps]{revtex4}

\usepackage{graphicx,amsmath,amssymb,color}

\begin{document}

\title{Determining the Quantum Expectation Value by Measuring a Single Photon}

\author{
F. Piacentini,$^{1\ast}$  A. Avella,$^{1}$ E. Rebufello,$^{1,2}$ R. Lussana,$^{3}$ F. Villa,$^{3}$ A. Tosi,$^{3}$ M.~Gramegna,$^{1}$  G. Brida,$^{1}$ E. Cohen,$^{4}$ L. Vaidman,$^{5}$ I. P. Degiovanni,$^{1}$ M. Genovese$^{1}$}
\affiliation{$^{1}$INRIM, Strada delle Cacce 91, I-10135 Torino, Italy}
\affiliation{$^{2}$Politecnico di Torino, Corso Duca degli Abruzzi 24, I-10129 Torino, Italy}
\affiliation{$^{3}$Politecnico di Milano, Dipartimento di Elettronica, Informazione e Bioingegneria, Piazza Leonardo da Vinci 32, 20133 Milano, Italy}
\affiliation{$^{4}$H.H. Wills Physics Laboratory, University of Bristol, Tyndall Avenue, Bristol, BS8 1TL, U.K}
\affiliation{$^{5}$Raymond and Beverly Sackler School of Physics and Astronomy, Tel-Aviv University, Tel-Aviv 6997801, Israel}
\affiliation{$^\ast$To whom correspondence should be addressed; E-mail: f.piacentini@inrim.it}


\baselineskip24pt

\maketitle

\textbf{Quantum mechanics, one of the keystones of modern physics, exhibits several peculiar properties, differentiating it from classical mechanics. One of the most intriguing is that variables might not have definite values. A complete quantum description provides only probabilities for obtaining various eigenvalues of a quantum variable. These and corresponding probabilities specify the expectation value of a physical observable, which is known to be a statistical property of an ensemble of quantum systems. In contrast to this paradigm, we demonstrate a unique method allowing to measure the expectation value of a physical variable on a single particle, namely, the polarisation of a single protected photon. This is the first realisation of quantum protective measurements.}
%
%
%
\\\\
Quantum theory has led to an unprecedented success in predicting a vast amount of experimental results, with a perfect agreement between its predictions and every realised connected experiment.
However, until this day there is no consensus about the foundational concepts of quantum mechanics. The reality of the wavefunction is still under hot debate \cite{PBR,Hardy2013,Ring,ad}. Probably the most puzzling feature of quantum theory, standing in stark contrast with classical physics, is the fact that physical observables lack definite values. A complete description of a quantum system only predicts the spectrum and probabilities for the measurement outcomes of a physical observable. Given the quantum state of the system $|\Psi\rangle$, which, according to standard quantum mechanics, comprises its complete description, to each observable $A$  we can associate a definite number: $\langle\Psi|A|\Psi\rangle =\sum p_i a_i$ ($p_i$ being the probability to obtain the (eigen)value $a_i$ as the result of the measurement of $A$). The meaning of this number is statistical: for finding the expectation value of $A$ it is necessary to have an ensemble of identically prepared systems and to perform numerous measurements.

Single measurement yielding the expectation value of a physical variable seems to be against the spirit of quantum mechanics.
However, it has been suggested that, in certain special situations, one can find the expectation value of an observable performing only a single measurement. This is the method of {\it protective measurement} (PM), originally proposed as an argument supporting the reality of the quantum wavefunction \cite{Prot1}.

The conceptual strength of the PM argument for the reality of the quantum state is a highly controversial issue \cite{Rovelli,Unruh,d,Meaning,Dass,Uffink,Prot4}, namely, does the procedure allow observing the state or only the protection mechanism?
Nevertheless, the idea of PM triggered and helped studying various foundational topics beyond exploring the meaning of the wavefunction, such as the analysis of Bohmian trajectories \cite{Traj2}, determination of the stationary basis \cite{Prot4} and analysis of measurement optimisation for minimising the state disturbance \cite{Sch}.
The concept of protection was also extended to measurement of a two-state vector \cite{PPP}. In spite of the rich and diverse analysis of the theory behind PM, to this date PM have not been realised experimentally.

Protection can be realised \cite{Prot1} both actively or passively: in this paper we employ an active protection technique based on the Zeno effect \cite{Misra}. It is worth mentioning that, although weak measurements (WMs) \cite{wm} and Zeno effect \cite{z1,z2,z3,z4,z5,z6} have been largely considered in experiments for several physical systems, up to now no experiment joining them in a PM has been realized yet.

\begin{figure}[tbp]
\begin{center}
\includegraphics[width=0.69\columnwidth]{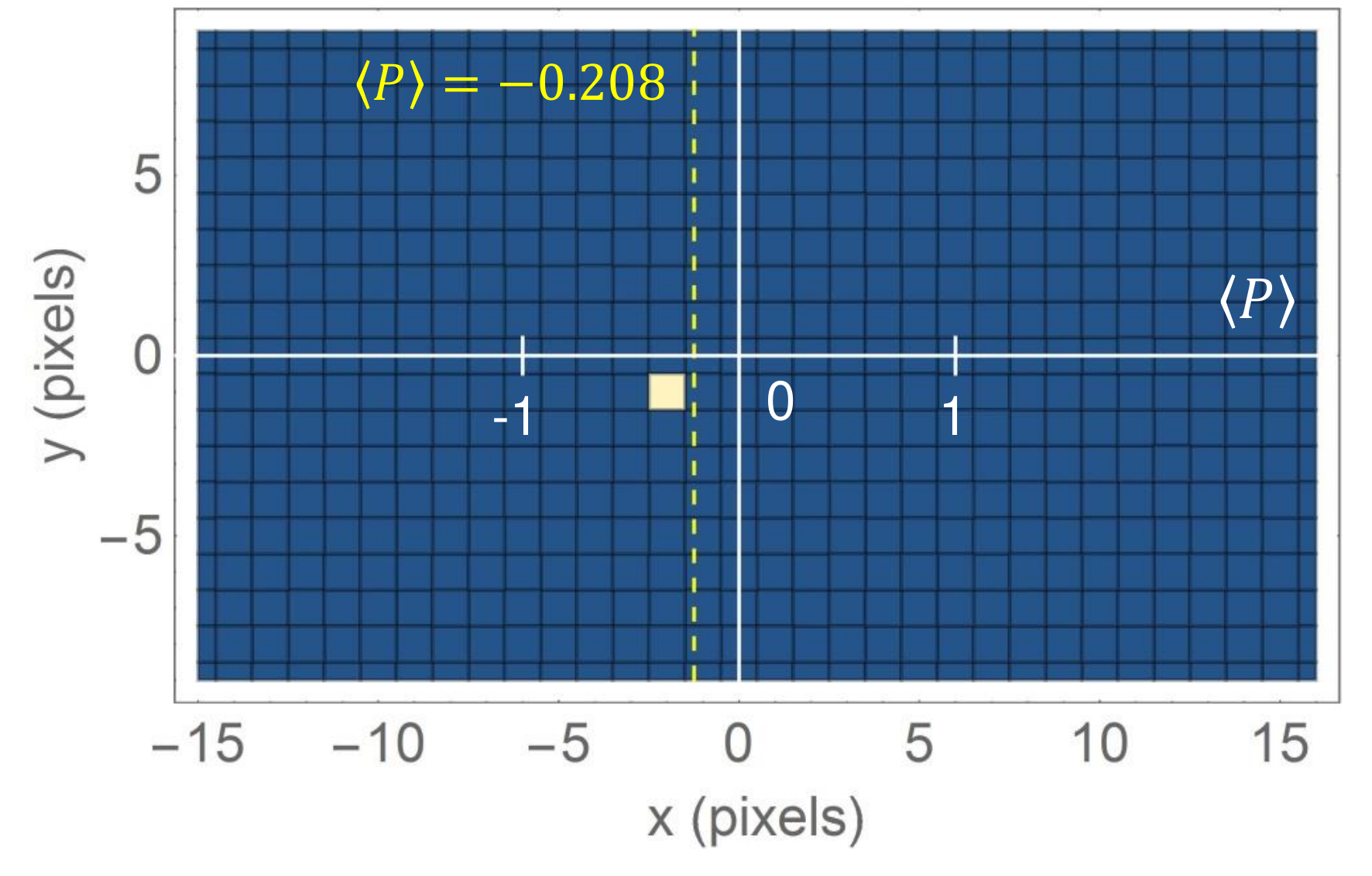}
\caption{\textbf{Estimation of the polarisation expectation value $\langle P \rangle$ by means of a single protective measurement.}
The $x$ coordinate of the pixel which detected the single photon tells us - without the need of any statistics - the expectation value of the polarisation operator,  $\langle P \rangle= -0.3(3)$, where the uncertainty is estimated from the width of the photon counts distribution presented in the paper, the theoretical value being (for $\theta =\frac{17\pi}{60}$) $\langle P \rangle= -0.208$.
} \label{claim}
\end{center}
\end{figure}

Our main result is the extraction of the expectation value of the photon polarisation by means of a measurement performed on a single protected photon (see Fig. \ref{claim}), that survived the Zeno-type protection scheme. The polarisation operator is defined by
\begin{equation}
P = |H\rangle\langle H|-|V\rangle\langle V|,
\end{equation}
where $H$ and $V$ are the horizontal and vertical polarisations, respectively. We note that, because of the presence of the active protection in our experiment, the single click of a multi-pixel camera tells us that the expectation value of the polarisation operator of the single protected photon is $\langle P \rangle= -0.3\pm 0.3$ (see Fig. \ref{claim}), in agreement with the theoretical predictions ($\langle P \rangle= -0.208$).

\begin{figure}[htbp]
\begin{center}
\includegraphics[width=0.89\columnwidth]{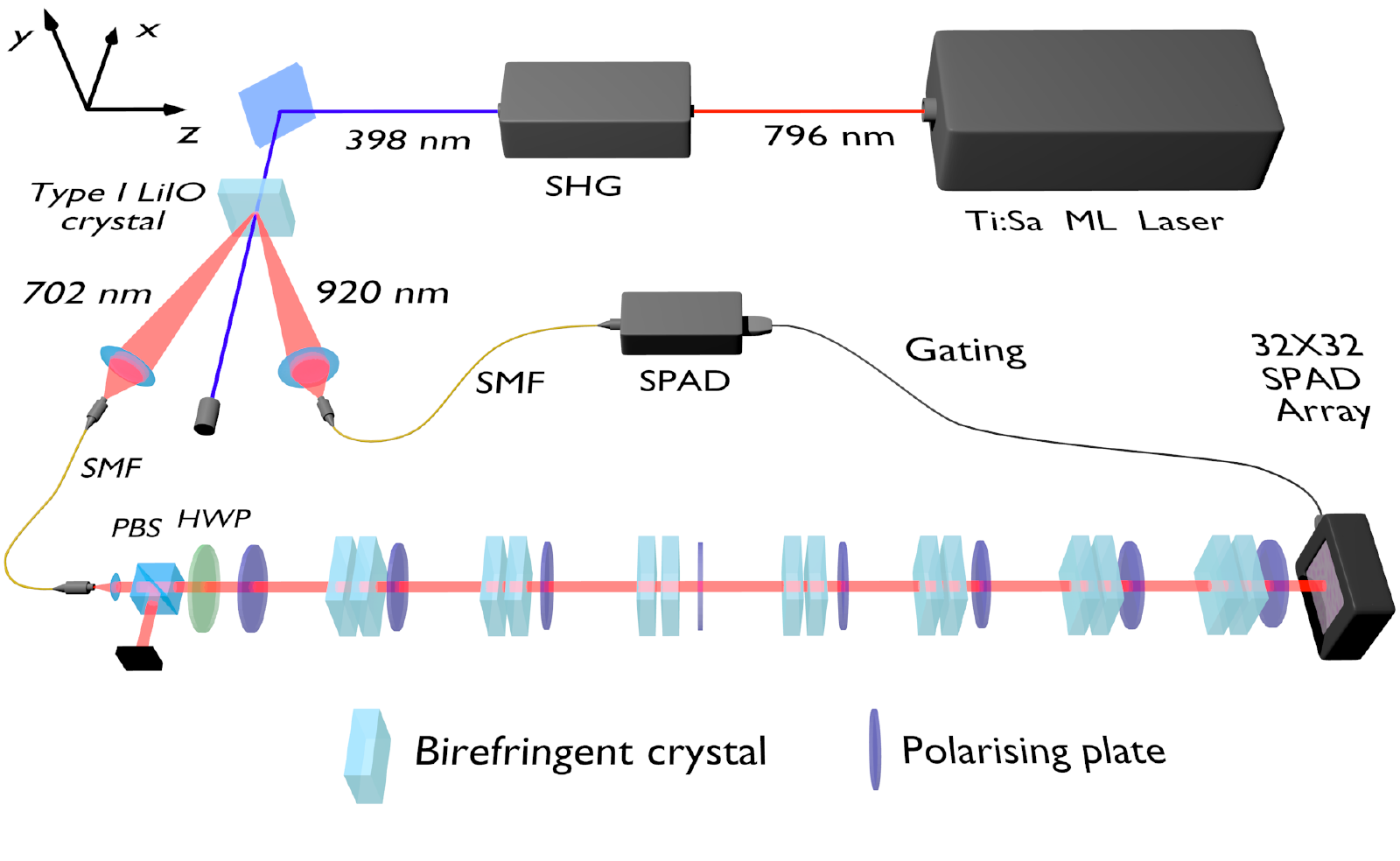}
\caption{\textbf{Experimental setup.} Heralded single photons are produced by type-I Parametric Down-Conversion in a LiIO$_3$ crystal, then properly filtered, fiber coupled and addressed to the open-air path where the experiment takes place.
After being prepared in the polarisation state $| \psi_{\theta}\rangle = \cos\theta | H \rangle + \sin\theta | V \rangle$, they pass through a birefringent material shifting them in the transverse direction $x$ (according to their polarisation).
The weak interaction is obtained by means of $K=7$ birefringent units, each unit composed of a first crystal separating the beam by 1.66 pixels (less than the beam width) and a second one used to compensate the phase and time decoherence induced by the first crystal: only the action of all units together allows separating orthogonal polarisations.
The protection of the quantum state, implementing the quantum Zeno scheme, is realised by inserting a thin-film polariser after each birefringent unit, projecting the photons onto the same polarisation as the initial state.
At the end of the optical path, the photons are detected by a spatial-resolving single-photon detector prototype, i.e., a two-dimensional array of $32\times32$ ``smart pixels''.
}
\label{setup}
\end{center}
\end{figure}

\begin{figure}[tbp]
\begin{center}
\includegraphics[width=0.99\columnwidth]{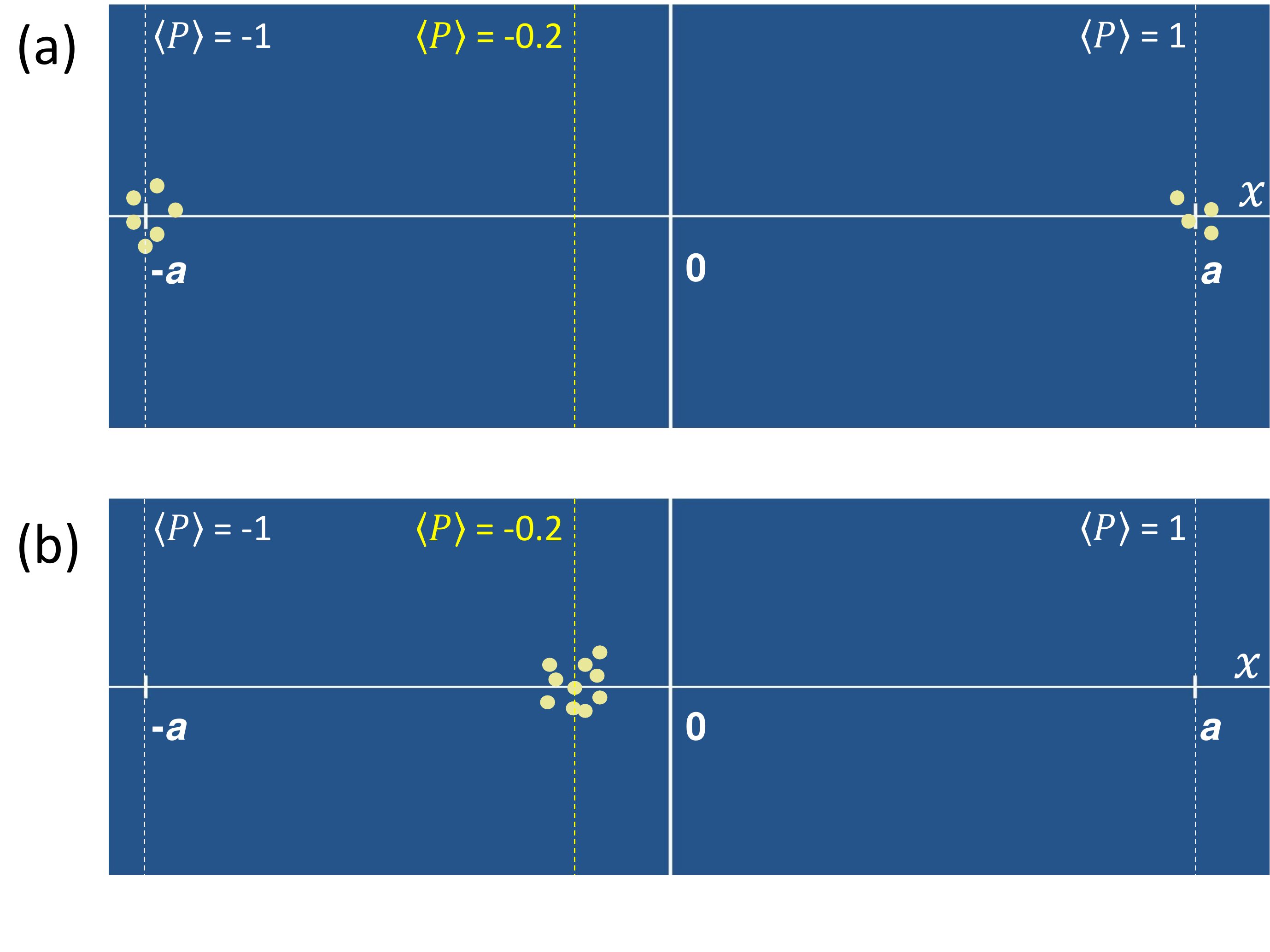}
\caption{\textbf{Illustrative drawing showing the measurement of unprotected (panel (a)) and protected (panel (b)) photons.} In the first case, $6$ photons fall close to $x=-a$ (corresponding to $P = -1$), while $4$ photons fall near $x=a$ (corresponding to $P = 1$), giving the expectation value $\langle P \rangle = -0.2$. In the second case, instead, all the photons accumulate close to $x=-0.2a$, the $\langle P \rangle = -0.2$ position; this indicates that, with PM, we can estimate the expected value of our observable even with a single photon.}
\label{gedanken}
\end{center}
\end{figure}

Our experiment (see Fig. \ref{setup}) is analogous to a Stern-Gerlach experiment.
Heralded single photons, prepared in the polarisation state $| \psi_{\theta}\rangle = \cos\theta | H \rangle + \sin\theta | V \rangle$, pass through a birefringent material shifting them in the transverse direction $x$ (according to their polarisation).
The spatial mode is close to a Gaussian one with a 4.1 pixels $\sigma$ (being $\sigma$ the source of uncertainty associated with the estimation of $\langle P \rangle$ presented in Fig. \ref{claim}).
The WM interaction is obtained exploiting $K=7$ birefringent units, while the state protection is realised via the quantum Zeno scheme (see Methods for experimental details).
At the end of the optical path, the photons are detected by a spatial-resolving single-photon detector prototype \cite{v}.
Without protection, the photons end up in one of the two regions corresponding to the vertical and horizontal polarisations, centered around $x=\pm a$ (see Fig. \ref{gedanken}a).
Then, the expectation value can be statistically found by counting the ratio of counts:
\begin{equation}\label{stat}
\langle P \rangle= \frac{N_H - N_V}{N}.
\end{equation}
In contrast, when employing PM, the photons end up in a region whose center is located at  $x=  a \langle P \rangle $ (see Fig. \ref{gedanken}b).
A large ensemble of measurements allows finding the center with arbitrarily good precision, but even a single photon detection provides information about $\langle P \rangle$, albeit with a finite precision defined by the width of the distribution.

\begin{figure}[tbp]
\begin{center}
\includegraphics[width=0.79\columnwidth]{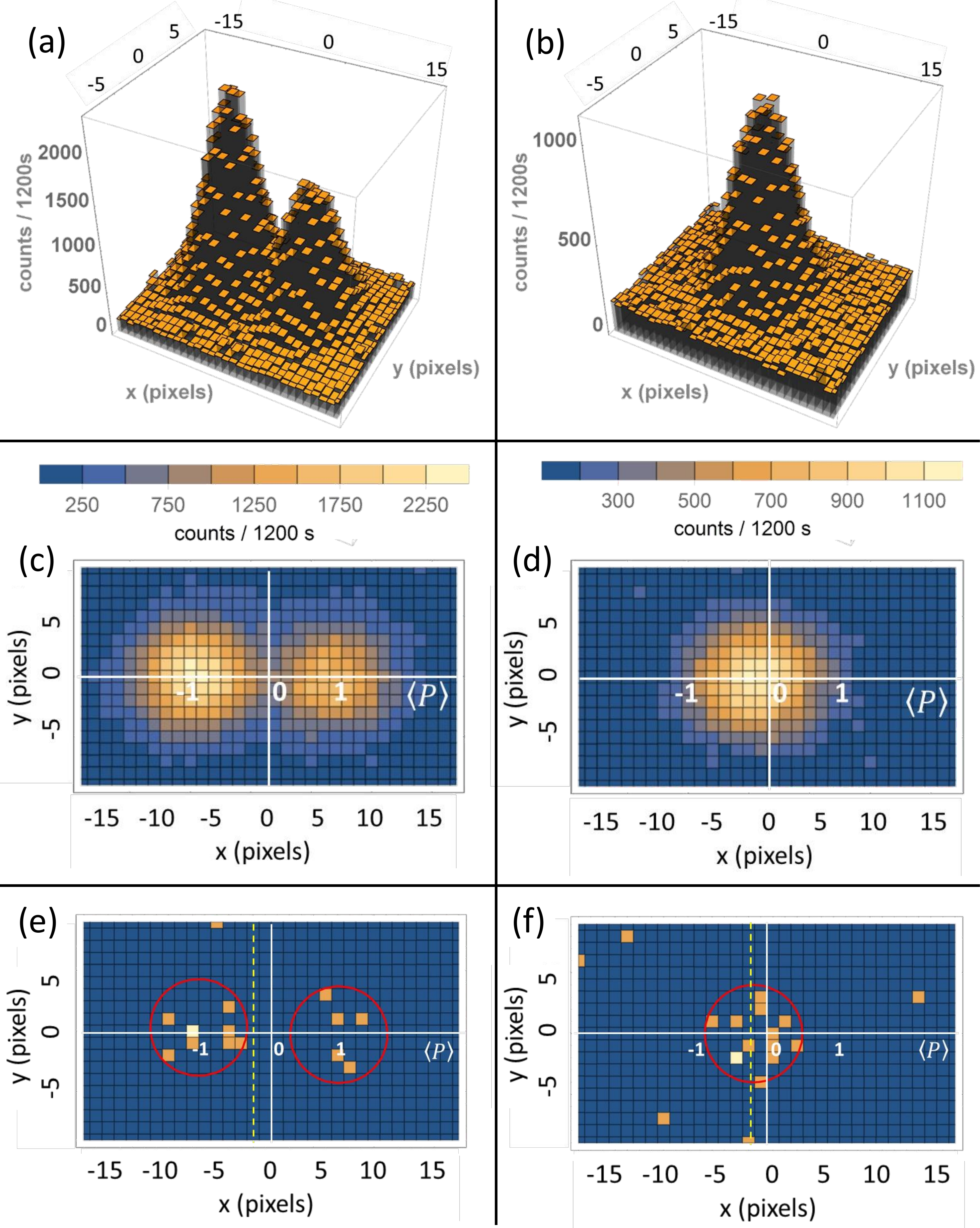}
\caption{\textbf{Results obtained for the input state $\mathbf{|\psi_{\frac{17\pi}{60}}\rangle=0.629|H\rangle+0.777|V\rangle}$.}
Panels (a) and (c): histogram and contour plot of the photon counts distribution obtained for the unprotected state.
Panels (b) and (d): histogram and contour plot of the photon counts distribution obtained for the protected state.
Panel (e): experiment with 14 single events (the first one in white), without protection. The yellow dashed line indicates the $x=a\langle P_\frac{17\pi}{60} \rangle$ value.
Panel (f): experiment with 17 single events (the first one in white), with protection: as expected, all the photons accumulate around the $x=a\langle P_\frac{17\pi}{60} \rangle$ position (yellow dashed line).
} \label{results1}
\end{center}
\end{figure}

In Fig. \ref{results1}(a-d) we show the results obtained collecting heralded single photons for a measurement time of 1200 s.
In panels (a) and (c) we see, respectively, a histogram and a contour plot of the photon counts distribution observed in the unprotected case for the input state $|\psi_{\frac{17\pi}{60}}\rangle =0.629|H\rangle+0.777|V\rangle$.
As in a standard Stern-Gerlach experiment, we observed photons only in two regions corresponding to the eigenvalues of $P$.
The expectation value of the  polarisation $\langle P \rangle$ evaluated using (\ref{stat}) from this distribution (after dark count subtraction) is $\langle P_\frac{17\pi}{60} \rangle = -0.21(4)$, in  agreement with theoretical expectations, $\langle P_\frac{17\pi}{60} \rangle = -0.208$.
Panels (b) and (d) show histogram and contour plot of the photon counts distribution obtained in the protected case for the same polarisation state.
Instead of two distributions around $x=\pm a$, here we find a single distribution of photon detections centered very close to the expected value $x = \langle P \rangle a$. The measured expectation value is $\langle P_\frac{17\pi}{60} \rangle = -0.19(2)$ (dark counts subtracted).
This result demonstrates that we have been able to realise and exploit the PM concept, providing the estimation of the polarisation operator, $\langle P \rangle $, by the detection of a single photon.

This is further confirmed in Fig. \ref{results1} (e) and (f), presenting typical photon detection maps for the input state $| \psi_{\frac{17\pi}{60}}\rangle$ obtained from a small number of detected photons.
Specifically, Fig. \ref{results1}(e) and (f) correspond respectively to the unprotected case ($N=14$ detection events) and to the protected one ($N=17$ detection events); the circles drawn in the two figures represent the width of the distributions reported in Fig. \ref{results1}(a-d).
As expected, there is a clear concentration of the counts inside the circles that - despite the non-ideality of our SPAD array, presenting a non-negligible level of dark counts likely responsible for the detection events outside the circles- demonstrates the validity of our technique even when just few detections are considered. The first detected photons in the runs are signified with white pixels. We see that while the white pixel of Fig. \ref{results1}(f) provides a good estimate of the expectation value, we cannot learn much from the white pixel of Fig. \ref{results1}(e).

%
%

Indeed, using a single photon is what makes PM special.
Nevertheless, one could argue that our experiment concerns a single {\it post-selected} photon (i.e. that survived all protection stages) and allowing post-selection, one can perform a measurement yielding the expectation value in the case of both weak and strong interaction.

\begin{figure}[tbp]
\begin{center}
\includegraphics[width=0.99\columnwidth]{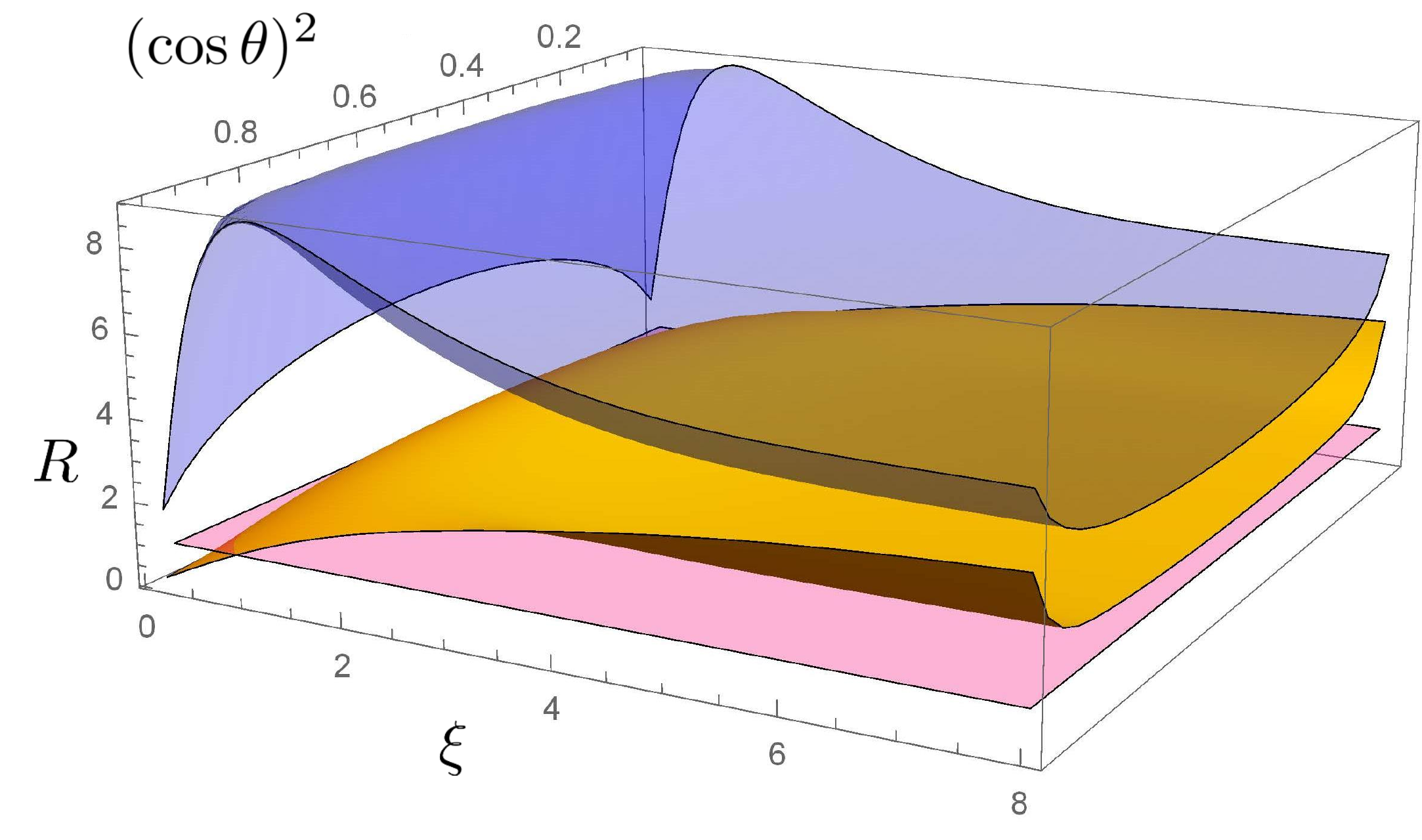}
\caption{\textbf{Comparison between the uncertainty on $\mathbf{P}$ with the PM approach ($\mathbf{u(P)}$) and the one given by projective measurement ($\mathbf{u_{\mathrm{PBS}}(P)}$).} Yellow surface: ratio $R= \frac{u_{\mathrm{PBS}} (P)}{u(P)} $ for a PM scheme with $K=7$ interaction-protection stages (as in our experiment), plotted versus the interaction strength $\xi$ and the $H$-polarisation component $(\cos \theta)^2$ of the single-photon state $| \psi \rangle$. Blue surface: ratio $R$ for a PM with $K=100$ stages. Magenta surface: $R=1$ bound, discriminating the part where PM approach is advantageous (above) and disadvantageous (below) with respect to the projective measurement.}
\label{R}
\end{center}
\end{figure}

It is obvious that the photon survival probability decreases when the interaction strength grows, while increasing the number of interaction-protection stages decreases the uncertainty, but also the survival probability.
To discuss quantitatively the performances of PM with respect to the straightforward alternative, a projective measurement exploiting, e.g., a polarising beam-splitter, we plot in Fig. \ref{R} the ratio $R= \frac{u_{\mathrm{PBS}} (P)}{u(P)}$  between the uncertainties on $\langle P \rangle$ ($u(P)$ and $u_{\mathrm{PBS}} (P)$  respectively, considering $u_{\mathrm{PBS}} (P)$ the uncertainty associated with the measurement procedure saturating the Quantum Cramer-Rao bound, i.e. being equal to the inverse of the square root of the Quantum Fisher information \cite{paris1}).
We consider in both cases the same initial resources (photon number), taking into account the photons lost along the interaction-protection steps (see Methods).
We consider two different scenarios:  $K=7$ (yellow surface) and $K=100$ (blue surface) interaction-protection stages.
One immediately notices that, in both cases, PM is almost always advantageous ($R>1$) with respect to the projective measurement, going below the $R=1$ plane (in magenta) only in presence of extremely weak interactions.
In our experiment, with $\xi\sim0.4$ and just $K=7$, a $10\%$ advantage is already present for most of the possible states, even if the maximum for $R$ corresponds to $\xi\sim1$.
For $K=100$, instead, the reasonably weak interaction $\xi\sim0.4$ grants the maximum of the advantage ($R>8.5$ almost everywhere), while for stronger interaction the advantage is reduced to $R<4$.
The advantage of the protective measurement technique stems from the very high survival probability of the protected photons.
This comes from the fact that, in presence of a sequence of identical interaction-protection stages as in our scheme, the relative probability of losing a photon in a protection step decreases with the single photon advancing in the sequence, since it is more likely to find the photon close to the ``right paths'' created in our $K$ interaction-protection steps (see Methods for theoretical details).

It is also worth pointing out that our experiment is the first realisation of a ``robust'' WM  \cite{preAAV} at single photon level.

To conclude, our results demonstrate that a single-event detection can provide a reliable information regarding a certain property of a quantum system -the expectation value of the polarisation operator in our case- which was considered to be only statistical, belonging to an ensemble of identically prepared quantum system.
This means that PM can be useful in practical situations where one wants to test an unknown state preparation procedure, taking advantage of the fact that both the state preparation and the state protection are performed exploiting the same projective measurement system (or equivalently a set of identical projective measurements as in our case).
This is the first experimental realisation of protective measurements \cite{Prot1}.

\section*{Acknowledgements}
This work has been supported by EMPIR-14IND05
``MIQC2'', and by the John Templeton Foundation (Grant No. 43467). E.C. was supported by ERC AdG NLST. L.V. acknowledges support of the Israel Science Foundation Grant No. 1311/14 and the German-Israeli Foundation for Scientific Research and Development Grant No. I-1275-303.14

\section*{Competing financial interests}

The authors declare no competing financial interests.\\\\

\section{Methods}

\subsection{Setup}

Our experimental setup (Fig. \ref{setup}) is composed of three parts.
In the first one -the preparation stage- we produce single photons in well-defined polarisation states, by means of a heralded single-photon source (SPS) based on Type-I Parametric Down-Conversion (PDC) \cite{hs}, and a state filtering system.
The spatial mode is close to a Gaussian of width 4.1 pixels.

The second part contains a sequence of weak interactions and state protection mechanism, the latter based on polarisation filters.

Finally, the photons are detected by a single-photon detector with a spatial resolution consisting of an array of single-photon detectors.

Incidentally, the scheme for protective measurement has got some analogy with to the one realised in \cite{Piacentini2016,Thekkadath2016}, but of course these two papers, aimed at realising sequential weak measurements, do not consider any protection mechanism. Furthermore, in these cases two weak interactions were considered, while in the actual scheme we have realised a challenging sequence of 7 interactions in a row.

%
%
%
The SPS is based on a 796 nm mode-locked Ti:Sapphire laser (repetition rate: 76 MHz), whose second harmonic emission pumps a $10 \times 10 \times 5$ mm LiIO$_3$ non-linear crystal, in which correlated photons are produced by PDC.
The idler photons ($\lambda_i = 920$ nm) are coupled to a single-mode fiber (SMF) and then addressed to a Silicon Single-Photon Avalanche Diode (SPAD), heralding the presence of the correlated signal photons ($\lambda_s = 702$ nm). These, after being SMF-coupled, are addressed to a launcher injecting them into the free-space optical path, where the protective measurement protocol is implemented.
After the launcher, the heralded single photons are collimated by a telescopic system, and then prepared in the linear polarisation state $|\psi_\theta \rangle$ (by means of a calcite polariser followed by a half-wave plate and a polariser).
We have estimated the quality of our single-photon emission with a Hanbury-Brown and Twiss interferometer, obtaining a value for the parameter $\alpha$ \cite{grangier} (directly connected to the second-order Glauber autocorrelation function $g^{(2)}$) of $0.13 \pm 0.01$ without any background or dark count subtraction, that being largely below 1 testifies the quality of our single photon source.

The Hamiltonian evolution of the quantum state is induced by exploiting  birefringence.
In our optical path we can insert up to $K=7$ birefringent units, each of them composed of two different calcite crystals  (the number $K=7$ of verification measurements was chosen from practical considerations approximating the ideal case of large $K$; because of losses originating from optical elements imperfections and because of the detector non-unit quantum efficiency and dark counts, further increasing $K$ would result in a low photon survival probability, and consequently a low signal-to-noise ratio at the detector output).
The first crystal of each element is a $2$ mm long birefringent crystal whose extraordinary (e) optical axis lies in the Y-Z plane, with an angle of $\pi/4$ with respect to the $z$ direction.
Due to the spatial walk-off effect experienced by the horizontally-polarised photons (i.e. along the $x$ direction), horizontal and vertical-polarisation paths get slightly separated along the $x$ direction (coupling to the pointer variable).
The second crystal of each unit is a $1.1$ mm long birefringent crystal with the optical e-axis orthogonal to the beam (thus not contributing to the spatial walk-off) that nullifies, through phase compensation, the temporal walk-off introduced by the first one.

The protection of the quantum state, implementing the quantum Zeno scheme, is realised by inserting a thin-film polariser after each birefringent unit, projecting the photons on the same polarisation of the initial state $|\psi_\theta \rangle$. Uhlmann's fidelities between reconstructed states and theoretical expected states always exceed $99\%$.
%

At the end of the optical path, the photons are detected by a spatial-resolving single-photon detector prototype.
This device is a two-dimensional array made of $32\times32$ ``smart pixels'' - each hosting a Silicon Single-Photon Avalanche Diode (SPAD) detector and its front-end electronics \cite{v}, but we used only $32\times20$ pixels to avoid distortion due to dark counts in the regions with negligible photon detection probability).
The SPAD array is gated with a 6 ns integration windows, triggered by the SPAD in the heralding arm in order to reduce the dark counts and improve the signal-to-noise ratio.

Aside from the single-photon state $|\psi_{\frac{17\pi}{60}}\rangle=0.629|H\rangle+0.777|V\rangle$, with which we obtained the data set presented in the paper, we tested our setup with other two states, namely $|\psi_{\frac{\pi}{4}}\rangle=\frac{1}{\sqrt{2}}\left(|H\rangle+|V\rangle\right)$ and $| \psi_{\frac{\pi}{8}}\rangle = 0.924|H\rangle+0.383|V\rangle$.

\begin{figure}[ht]
\begin{center}
\includegraphics[width=0.99\columnwidth]{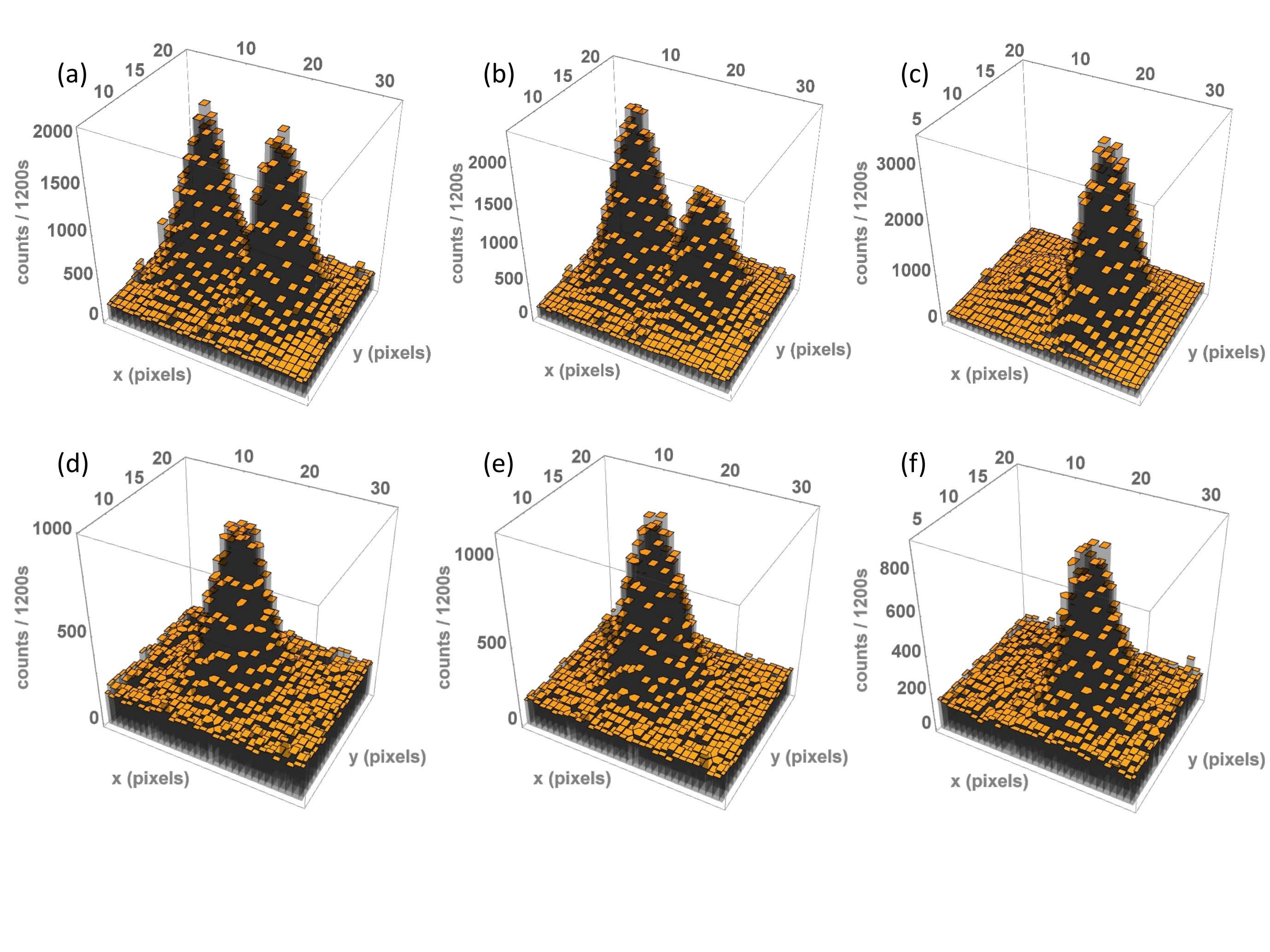}
\caption{\textbf{Results obtained with the different polarisation input states exploited.} Plots (a), (b) and (c): photon count distributions obtained with three different input states ($| \psi_{\frac{\pi}{4}}\rangle =\frac{1}{\sqrt{2}}(|H\rangle+|V\rangle)$, $| \psi_{\frac{17\pi}{60}}\rangle =0.629|H\rangle+0.777|V\rangle$ and $| \psi_{\frac{\pi}{8}}\rangle = 0.924|H\rangle+0.383|V\rangle$ respectively) in 1200 s, without protection. In all these plots two different peaks centered in $x=\pm a$, corresponding to horizontally- and vertically-polarised single photons, are present.
Plots (d), (e) and (f): photon counts distributions obtained with the same set of states in 1200 s, protected. In contrast to what present in the three panels above, here we can see a single peak of photon counts, centered in correspondence of $x=\langle P\rangle a$, demonstrating how, with protective measurements, each photon carries information about the expectation value $\langle P\rangle$.
In both the protected and unprotected cases, dark counts are estimated and subtracted before evaluating $\langle P \rangle$.
} \label{f3_results}
\end{center}
\end{figure}

The obtained results for these data sets, together with the one shown in the main paper, are summarized in Fig. \ref{f3_results}.
As immediate to see, in the unprotected case the photon counts accumulate around the positions $x=\pm a$, forming the distributions reported in panels (a), (b) and (c) for the states $|\psi_{\frac{\pi}{4}}\rangle$, $|\psi_{\frac{17\pi}{60}}\rangle$ and $|\psi_{\frac{\pi}{8}}\rangle$, respectively.
From these photon-counts distributions, we can evaluate the expectation value of each state polarisation, obtaining $\langle P_\frac{\pi}{4} \rangle = -0.03(4)$, $\langle P_\frac{17\pi}{60} \rangle = -0.21(2)$, $\langle P_\frac{\pi}{8} \rangle = 0.72(2)$, all in excellent agreement with the theoretical expectations ($\langle P_\frac{\pi}{4}\rangle=0$, $\langle P_\frac{17\pi}{60} \rangle = -0.208$ and $\langle P_\frac{\pi}{8} \rangle = 0.707$, respectively).

When we introduce the protection mechanism, instead, we notice (panels (d), (e) and (f)) that all the photons accumulate around a specific position, corresponding to $x=a\langle P \rangle$: as stated in the main text, this means that each single photon carries information about the expectation value of its polarisation, granting the possibility of extracting such value even in a one-shot experiment with just a single photon. The expectation values obtained in the protected case are respectively $\langle P_\frac{\pi}{4} \rangle = -0.012(14)$, $\langle P_\frac{17\pi}{60} \rangle = -0.19(2)$ and $\langle P_\frac{\pi}{8} \rangle = 0.72(2)$, in good agreement with the theory, as well as with the ones obtained without protecting the single photons.

In order to obtain these values, the dark counts of the $32\times32$ SPAD array, responsible for the ``floor'' of counts in all the six plots, were properly evaluated and subtracted.

\subsection{Theoretical analysis}

In the following we will describe the theory behind the protective measurement technique implemented in our experiment to extract the expectation value of the photon polarisation $P=|H\rangle \langle H | - |V\rangle \langle V | $ by means of a measurement performed on a single protected photon, where H and V are the horizontal and vertical polarisations. As explained in the paper, for doing this we take advantage of a spatial resolving detector, thus it is quite obvious to consider the space and polarisation degrees of freedom when we describe the wavefunction of our single photon, i.e. $|\Psi_{\mathrm{in}} \rangle  = |\psi \rangle \otimes | f_x \rangle$ with $ |\psi\rangle = \cos \theta | H \rangle + \sin \theta | V \rangle$ and $|f_x \rangle = \int \mathrm{d}x ~ f(x) |x \rangle $ where $f(x)$ is a Gaussian curve whose square is normalised to one, namely $f(x)=(2 \pi \sigma^2)^{-\frac{1}{4}} \exp(-\frac{x^2}{4 \sigma^2})$.

The protective measurement consists of a sequence of identical interactions, exploiting the spatial walk-off in non-linear crystals (using a technique completely analogous to the one used in, e.g., \cite{Piacentini2016}) corresponding to the unitary transformation $U=\exp(i g \mathbf{P}\otimes \Pi_{H} )$ (being $\mathbf{P}$ the momentum operator and $g$ the von Neumann coupling constant between $\mathbf{P}$ and the projector $\Pi_H=|H\rangle \langle H|$), and the protective measurement, performed by exploiting the polarisation projector $\Pi_{\psi}= |\psi \rangle \langle \psi |$.
Actually, this can be also described as a test on an unknown state preparation procedure exploiting a series of identical projective measurements, the first one really used to prepare (select) the single-photon state, while the other ones used as protective measurements.
After $K$ steps of this sequence consisting of a weak interaction followed by a protective measurement, the non-normalised output state of the single photons will be:
\begin{equation}
|\Psi_{\mathrm{out}} \rangle  = \left( \Pi_{\theta} U   \right)^{K} |\Psi_{\mathrm{in}} \rangle  = \left( \langle H| \Pi_{\theta} |H\rangle e^{ i g \mathbf{P}  } + \langle V| \Pi_{\theta} |V\rangle \mathbf{1}_x  \right)^{K}  |\Psi_{\mathrm{in}}\rangle.
\end{equation}

Thus, the probability that the protected single-photon survives after $K$ interaction-verification steps is $p_{\mathrm{sur}} (K) = \mathrm{Tr}[|\Psi_{\mathrm{out}} \rangle  \langle \Psi_{\mathrm{out}} | ] $, while the probability of finding the protected single-photon in a specific position $x_0$ of our spatial-resolving detector (given that it survived the verification process) is
\begin{eqnarray}
F_K (x_0)&=& p_{\mathrm{sur}} (K)^{-1} ~ \mathrm{Tr}[|x_0\rangle \langle x_0 |\Psi_{\mathrm{out}} \rangle  \langle \Psi_{\mathrm{out}} | ]
\\
&=& p_{\mathrm{sur}} (K)^{-1} ~ \left(\sum_{n=0}^{K} \frac{K!}{n! (K-n)!} \langle H| \Pi_{\theta} |H\rangle^{n} \langle V | \Pi_{\theta} |V \rangle^{K-n} f(x_0 + n g)    \right)^2.  \label{FNx}
\end{eqnarray}

As explained in the paper, the spatially-resolved detection of the protected single-photon provides an estimation of the value of $P$.
A relevant question is related to the quality of this estimation, i.e. the uncertainty that can be associated with this estimation.
This uncertainty is obviously associated to the uncertainty on the arrival position of one protected single photon, related to the probability distribution profile $F_K (x)$.
The uncertainty in the position is $u(x)=\sqrt{ \epsilon(x^2) - \epsilon(x)^2 }$, with $\epsilon(x^n)=\int \mathrm{d}x ~x^n F_K (x) $.
Then, we note that there ia a correspondence between the value of $\langle P \rangle$ and the average position $\epsilon(x)$, where the protected single photon is detected, i.e., the $H$-polarisation corresponds to $\langle P \rangle=1$ and the position $\epsilon(x)=\frac{K g}{2}$, while the $V$-polarisation corresponds to $\langle P \rangle=-1$ and the position $\epsilon(x)=-\frac{K g}{2}$ (actually, the $V$-polarisation is not affected by the interaction according to the unitary interaction operator $U$, while the $H$-polarisation is shifted by $K g$, because of the fact that each interaction induces a shift $g$, according to Eq. (\ref{FNx}). Anyway, we decided to shift the whole $X$ axis by $-\frac{K g}{2}$ for an easier comparison with the well-known Stern-Gerlach experiment). Thus, the uncertainty on $\langle P \rangle$ associated with the detection of a single protected photon can be obtained simply by re-scaling the spatial uncertainty $u(x)$, i.e. $u(P)=  u(x)~ \frac{2}{K g } $.

It is interesting to compare the performance of the protective measurement technique presented above in situations of strong and weak interaction, corresponding to $g \gg \sigma $ and $g \ll \sigma $ respectively, versus the usual technique involving a single strong measurement, for example exploiting a polarizing-beam-splitter (PBS). In the PBS measurement, starting from $M$ initial photons in the polarisation state $| \psi \rangle$ the probability distribution of observing $m$ photons $V$-polarised (and, obviously, $(M-m)$ $H$-polarised) is the binomial-one with probability parameter $(\cos \theta)^2$. Thus, the estimator of $P$ is $ P  = \frac{2 m }{M} -1$ and the uncertainty on this estimator can be evaluated as $u_{\mathrm{PBS}} (P)= \sqrt{\langle P^{2} \rangle - \langle P \rangle^{2}}= \frac{|\sin(2 \theta)|}{\sqrt{M}} $. This uncertainty level is easily demonstrated to be the optimal one in terms of quantum Fisher information (indeed the quantum Fisher Information is $\mathcal{H}= \sin(2 \theta)^{-2} $), i.e. the one that saturates the quantum Cramer-Rao bound \cite{paris1}.

In order to provide a fair comparison between this PBS-based measurement and our protective-measurement-based one we should consider the two measurement approaches exploiting the same initial resources, i.e. the same number of initial prepared photons. In the protective measurement case we have considered the uncertainty associated with the detection of a protected single-photon, but the probability of survival of a $K$-step protective measurement process is $p_{\mathrm{sur}} (K)$, this means that to have one protected photon arriving at our ideal spatial resolving detector we need, on average, $1/p_{\mathrm{sur}} (K)$ initial photons.
To perform a fair comparison, we set $M=1/p_{\mathrm{sur}} (K)$ also in the PBS measurement case, and we define the ratio between the uncertainties $R= \frac{u_{\mathrm{PBS}} (P)}{u(P)} $.
This is what we show in Fig. 4 of the main text, where $R$ plotted versus the strength of the interaction ($\xi$ in the main text called interaction strength corresponds to the ratio $g/\sigma$) and the (linear) polarisation state of the single photon $| \psi \rangle$ (specifically, $(\cos \theta)^2$ is $\langle \psi | \Pi_H | \psi \rangle$), for $K=100$ and $K=7$ interactions. As already said, the protective measurement procedure is almost always advantageous ($R>1$) with respect to the PBS measurement, and it becomes disadvantageous only in the presence of extremely weak interactions (it is not shown in  the figure, but, e.g., also for $K=100$ and $g/\sigma = 0.02$ and $|\psi\rangle = 2^{-1/2}(|H\rangle+ |V\rangle)$ we have $R=0.996$).


In the case of $K=7$ ($K=100$) interactions, the maximum advantage is $R \sim 3$ ( $R \sim 8.5$) for $\xi \sim 1 $ ($\xi \sim 0.4 $), while a strong interaction reduces our factor to $R \sim 1.6$ ($R \sim 4$).
We underline that, considering a weak interaction exploiting, as in our experimental configuration, birefringent crystals separating the $H$- and $V$-photons paths of 1.66 pixels, but with $K=100$ interaction-protection stages, one would observe a total separation of 166 pixels between $H$ and $V$ polarisations, an order of magnitude above the FWHM of the single-photons spatial mode (a scenario similar to the one of Fig. 2 in the main text).

Here we want to stress that, even though projective measurement is an optimal measurement for the parameter $P$, i.e. a measurement procedure able to reach the quantum Cramer-Rao bound \cite{paris1}, protective measurement allows obtaining an estimation of $\langle P \rangle$ far better than the one achievable with such method, surpassing this way the quantum Cramer-Rao bound itself (being the state preparation procedure embedded in the protection scheme). $R$ can be understood also as a parameter identifying the resources needed to achieve the same level of uncertainties in the two measurement techniques: in fact, to achieve the same level of uncertainty, the initial number of photons needed in the PBS measurement is $R^2$, the one used in the protective measurement.

The advantage of the protective measurement technique comes from the very high survival probability of the protected photons. As evident from Fig. SM3, indeed, with $K=100$ interactions and  $g/\sigma \sim 0.4 $ we have $p_{\mathrm{sur}}>0.57$ even for the most lossy state. Also in the case of strong interaction $g/\sigma = 6 $ the survival probability is surprisingly high ($p_{\mathrm{sur}}>0.05$).

To understand this, one should consider the situation in which the protective measurement is performed without weak interactions: one has just to replace the ``weak'' birefringent crystals by ``strong'' ones, or to use a beam of narrow width, such that the shift due to each crystal will be much larger than the width of the beam (the situation depicted in Fig. \ref{f2_AddMat}). Then, the readings of the detector provide the expectation value with a precision scaling as $1/\sqrt K$, where $K$ is the number of interaction-protection stages.
The $1/\sqrt K$ uncertainty scaling with the number of interactions $K$ presents some analogy with the one of the projective measurement, where the uncertainty scales of a factor $1/\sqrt N$ with the number $N$ of exploited photons.

\begin{figure}[tbp]
\begin{center}
\includegraphics[width=0.99\columnwidth]{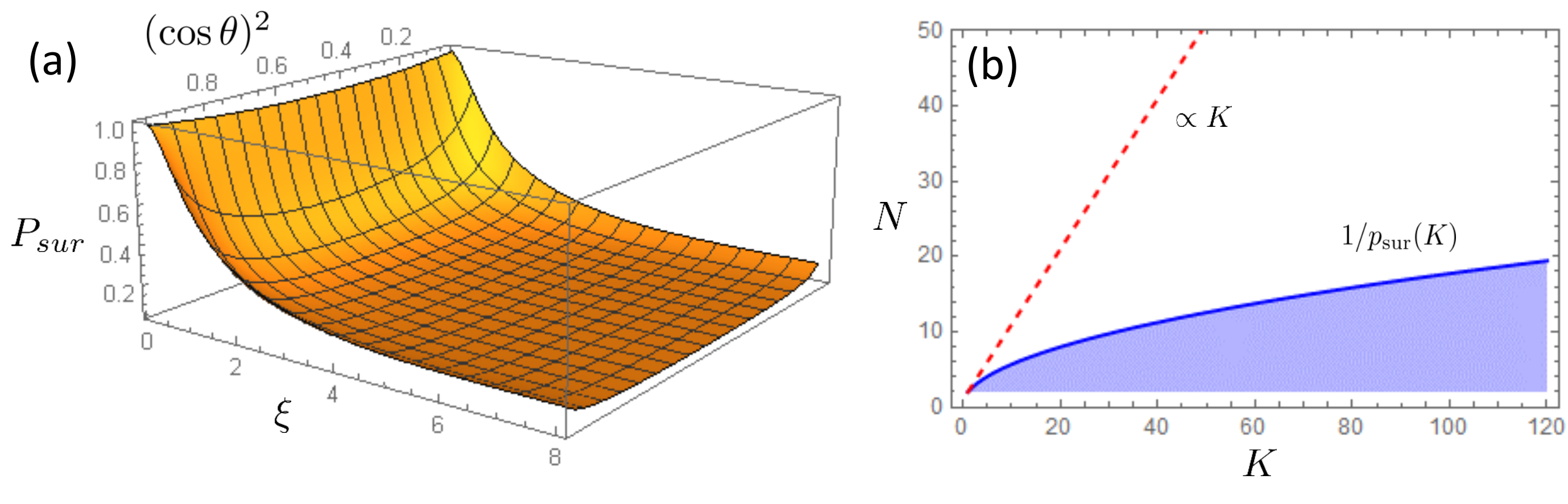}
\caption{\textbf{Photon survival probability in a protective measurement scheme.} Plot (a): Protected photon survival probability $p_{\mathrm{sur}}(K=100) $ plotted versus $\xi=g/\sigma$ (representing the interaction strength) and $(\cos \theta)^2$ (i.e. the $H$-polarisation component) of the single-photon state $| \psi \rangle$. Plot (b): the shaded area represents the average number $N$ of initial photons needed to observe a protected photon at the end of the protective measurement process ($1/p_{\mathrm{sur}} (K)$) versus the number of interaction-protection stages $K$ (the worst-case scenario, indicated by the blue curve on the boundary, corresponds to the state $|\psi_\frac{\pi}{4}\rangle$). For comparison, a dashed line corresponding to the linear scaling with $K$ is reported.}\label{P_sur}
\end{center}
\end{figure}

\begin{figure}[tbp]
\begin{center}
\includegraphics[width=0.99\columnwidth]{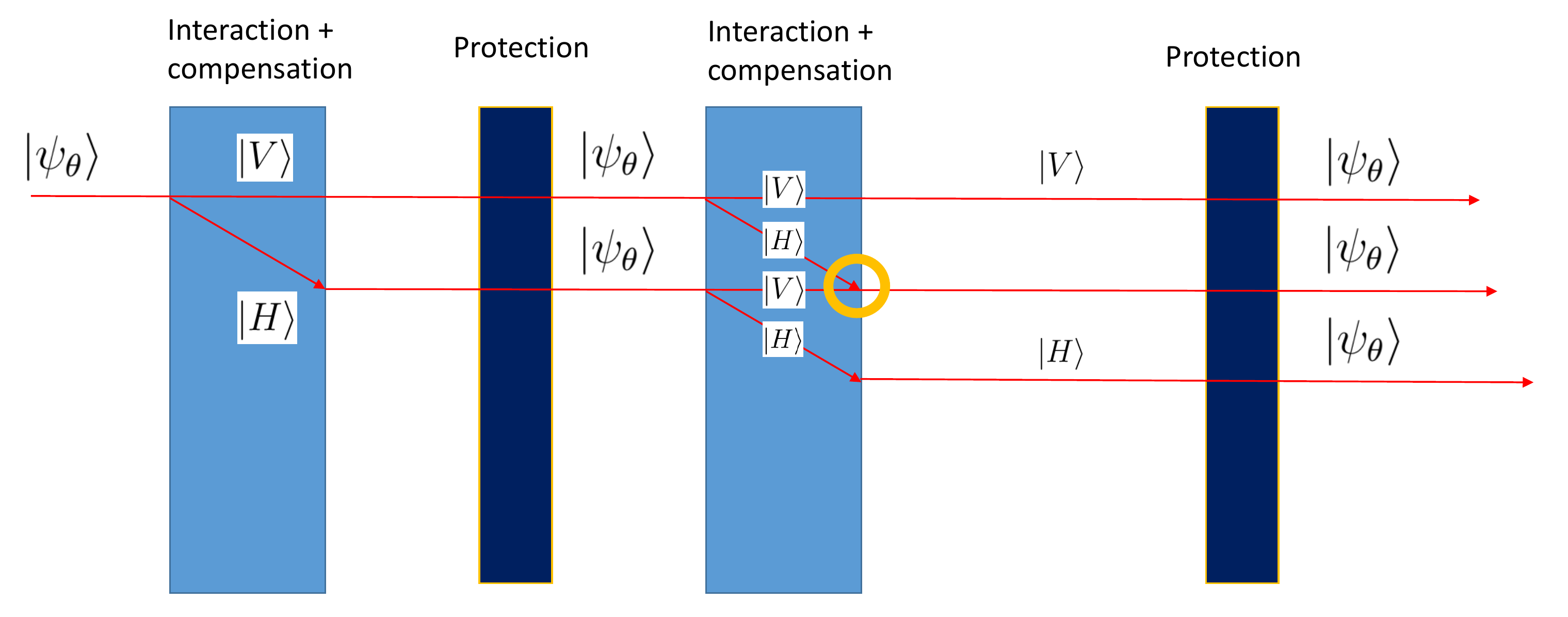}
\caption{\textbf{Protected measurement scheme in case of strong interaction $\mathbf{\xi>>1}$}. here, the red lines represent the photon beams, with the $H$- and $V$-polarisation completely separated by the birefringent crystals. From the second birefringent crystal pair (interaction+compensation blocks) onwards, the $H$- and $V$-beam recombine in some spots (indicated by the yellow circle) in a coherent way, forming a state with non-zero survival probability in the subsequent protection stage.
This effect results in a reduction of photon losses, specially close to the path corresponding to the expectation value of $\langle P\rangle$, granting advantage to the protective measurement scheme with respect to the projective one even without weak interaction (although lower than when weak coupling is exploited). This can be intuitively understood analising the probability distribution of the survived photons considering $|\psi_{\theta} \rangle = (| H \rangle +|V \rangle )/\sqrt{2} $. } \label{f2_AddMat}
\end{center}
\end{figure}

The advantage can be understood observing the fact that the average number of photons needed for observing a single protected photons, i.e. $1/p_{\mathrm{sur}} (K)$, grows at a much slower pace with respect to $K$, as shown in Fig. \ref{P_sur} (b).
This comes from the fact that, in presence of a sequence of identical interaction-protection stages as in our scheme, the relative probability of losing a photon in a protection step because of unsuccessful protection measurement decreases with the single photon advancing in the sequence since photons are more likely on the
``right path'' (see Fig. \ref{f2_AddMat}).


Obviously, the above considerations assume an ideal scenario where the the only source of losses in the protective measurement is represented by the projection measurement preforming the verification of the state, neglecting completely the optical losses induced by the real optical devices present in the experimental setup. In the specific case of our experiment, the optical losses greatly reduce the advantage discussed so far, but this is a technical limitation that can be, in principle, strongly reduced. It is important to underline that the advantage of the protective measurement technique can become absolutely relevant in a less lossy measurement scenario, and that our proof-of-principle experiment and simulations pave the way to the exploitation of the protective measurement approach in atomic or solid state quantum systems, where losses are typically less than the ones experienced in photon-based experiments.



\begin{thebibliography}{30}
%
%
\bibitem{PBR} M. F. Pusey, J. Barrett, T. Rudolph, On the reality of the quantum state. \emph{Nat. Phys.} \textbf{8}, 475-478 (2012).
%
\bibitem{Hardy2013} L. Hardy, Are quantum states real? \emph{Int. J. Mod. Phys. B} \textbf{27}, 1345012 (2013).
%
\bibitem{Ring} M. Ringbauer, B. Duffus, C. Branciard, E. G. Cavalcanti, A. G. White, A. Fedrizzi, Measurements on the reality of the wavefunction. \emph{Nat. Phys.} {\bf 11}, 249-254 (2015).
%
\bibitem{ad} M. Genovese, Interpretations of Quantum Mechanics and the measurement problem.  \emph{Adv. Sci. Lett.} 3, 249 (2010).
%
%
\bibitem{Prot1} Y. Aharonov, L. Vaidman, Measurement of the Schr\"{o}dinger Wave of a Single Particle. \emph{Phys. Lett. A} \textbf{178}, 38 (1993).
%
%
\bibitem{Rovelli} C. Rovelli, Comment on ``Meaning of the wave function''. \emph{Phys. Rev. A} \textbf{50}, 2788 (1994).
%
\bibitem{Unruh} W. G. Unruh, Reality and measurement of the wave function. \emph{Phys. Rev. A} \textbf{50}, 882 (1994).
%
%
\bibitem{d} G. M. D'Ariano, H. P. Yuen, Impossibility of measuring the wave function of a single quantum system. \emph{Phys. Rev. Lett.} \textbf{76}, 2832 (1996).
%
\bibitem{Meaning} Y. Aharonov, J. Anandan, L. Vaidman, The Meaning of Protective Measurements. \emph{Found. Phys.} \textbf{26}, 117-126 (1996).
%
\bibitem{Dass} N. H. Dass, T. Qureshi, Critique of protective measurements. \emph{Phys. Rev. A} \textbf{59}, 2590 (1999).
%
\bibitem{Uffink} J. Uffink, How to protect the interpretation of the wave function against protective measurements. \emph{Phys. Rev. A} \textbf{60}, 3474 (1999).
%
%
\bibitem{Prot4} S. Gao, Protective Measurement and Quantum Reality (Cambridge University Press, UK, 2015).
%
\bibitem{Traj2} Y. Aharonov, B. G. Englert, M. O. Scully, Protective measurements and Bohm trajectories. \emph{Phys. Lett. A} \textbf{263}, 137 (1999).
%
%
\bibitem{Sch} M. Schlosshauer, Measuring the quantum state of a single system with minimum state disturbance. \emph{Phys. Rev. A} \textbf{93}, 012115 (2016).
%
\bibitem{PPP} Y. Aharonov, L. Vaidman, Protective Measurements of Two-State Vectors. In \emph{Potentiality, Entanglement and Passion-at-a-Distance}, eds/ R. S. Cohen,
M. Horne and J. Stachel, BSPS 1-8, (Kluwer, 1997), quant-ph/9602009.
%
\bibitem{Misra} B. Misra, E. C. G. Sudarshan, The Zeno's paradox in quantum theory. \emph{J. Math. Phys.} \textbf{18}, 756 (1977).
%
\bibitem{wm} J. Dressel, M. Malik, F. M. Miatto, A. N.Jordan and R. W. Boyd, Understanding quantum weak values: Basics and applications. \emph{Rev. Mod. phys.} \textbf{86}, 307 (2014).

\bibitem{z1} W. M. Itano, D. J. Heinzen, J. J. Bollinger, and D. J. Wineland, Quantum Zeno effect. \emph{Phys. Rev. A} \textbf{41}, 2295 (1990).
\bibitem{z2} P. G. Kwiat, A. G. White, J. R. Mitchell, O. Nairz, G. Weihs, H. Weinfurter, A. Zeilinger, High-Efficiency Quantum Interrogation Measurements via the Quantum Zeno Effect. \emph{Phys. Rev. Lett.} \textbf{83}, 4725 (1999).
\bibitem{z3} J. M. Raimond, C. Sayrin, S. Gleyzes, I. Dotsenko, M. Brune, S. Haroche, P. Facchi and S. Pascazio, Phase Space Tweezers for Tailoring Cavity Fields by Quantum Zeno Dynamics. \emph{Phys. Rev. Lett.}\textbf{105}, 213601 (2010).
\bibitem{z4} L. Bretheau, P. Campagne-Ibarcq, E. Flurin, F. Mallet, B. Huard, Quantum dynamics of an electromagnetic
mode that cannot contain N photons. \emph{Science} \textbf{348}, 776 (2015).
\bibitem{z5} A. Signoles, A. Facon, D. Grosso, I. Dotsenko, S. Haroche, J. Raimond, M. Brune, S. Gleyzes, Confined quantum Zeno dynamics of a watched atomic arrow. \emph{Nat Phys.}\textbf{10}, 715 (2014).
\bibitem{z6} G. Mazzucchi, W. Kozlowski, S. F. Caballero-Benitez, T. J. Elliott and I. B. Mekhov, Quantum measurement-induced dynamics of many-body ultracold bosonic and fermionic systems in optical lattices. \emph{Phys. Rev. A} \textbf{93}, 023632 (2016).

\bibitem{v} F. Villa, R. Lussana, D. Bronzi, S. Tisa, A. Tosi, F. Zappa, A. Dalla Mora, D. Contini, D. Durini, S. Weyers, W. Brockherde,
CMOS Imager With 1024 SPADs and TDCs for Single-Photon Timing and 3-D Time-of-Flight. \emph{IEEE J. Sel. Top. Quantum Electron.} \textbf{20}, 
3804810 (2014).

\bibitem{paris1} M. G. A. Paris, Quantum Estimation For Quantum Technology. \emph{Int. J. Quantum Inform.} \textbf{07}, 125 (2009).
%
%
%
\bibitem{preAAV} Y. Aharonov, D. Z. Albert, A. Casher, L. Vaidman, Surprising quantum effects. \emph{Phys. Lett. A} \textbf{124}, 199-203 (1987).
%
%
%
%

\bibitem{hs}
G. Brida, I. P. Degiovanni, M. Genovese, F. Piacentini, P. Traina, A. Della Frera, A. Tosi, A. Bahgat Shehata, C. Scarcella, A. Gulinatti, M. Ghioni, S. V. Polyakov, A. Migdall, A. Giudice, An extremely low-noise heralded single-photon source: A breakthrough for quantum technologies. \emph{Applied Phys. Lett.} \textbf{101}, 221112 (2012) and ref.s therein.
%

%
\bibitem{Piacentini2016}
F. Piacentini, A. Avella, M. P. Levi, M. Gramegna, G. Brida, I. P. Degiovanni, E. Cohen, R. Lussana, F. Villa, A. Tosi, F. Zappa, and M. Genovese, Measuring Incompatible Observables by Exploiting Sequential Weak Values. \emph{Phys. Rev. Lett.} \textbf{117}, 170402 (2016).

\bibitem{Thekkadath2016} G. S. Thekkadath, L. Giner, Y. Chalich, M. J. Horton, J. Banker, and J. S. Lundeen, Direct Measurement of the Density Matrix of a Quantum System. \emph{Phys. Rev. Lett.} \textbf{117}, 120401 (2016).




\bibitem{grangier}
P. Grangier, G. Roger, A. Aspect, Experimental Evidence for a Photon Anticorrelation Effect on a Beam Splitter: A New Light on Single-Photon Interferences. \emph{Eur. Phys. Lett.} \textbf{1}, 173 (1986).



\end{thebibliography}
\end{document}